# Defining Estimands Using a Mix of Strategies to Handle Intercurrent Events in Clinical Trials

Yongming Qu*, Linda Shurzinske, Shanthi Sethuraman

Global Statistical Sciences, Eli Lilly and Company, Lilly Corporate Center

Indianapolis, IN 46285

*Corresponding email: qu_yongming@lilly.com

*June 4, 2020*

## Abstract

Randomized controlled trials (RCT) are the gold standard for evaluation of the efficacy and safety of investigational interventions. If every patient in an RCT were to adhere to the randomized treatment, one could simply analyze the complete data to infer the treatment effect. However, intercurrent events (ICEs) including the use of concomitant medication for unsatisfactory efficacy, treatment discontinuation due to adverse events, or lack of efficacy, may lead to interventions that deviate from the original treatment assignment. Therefore, defining the appropriate estimand (the appropriate parameter to be estimated) based on the primary objective of the study is critical prior to determining the statistical analysis method and analyzing the data. The International Council for Harmonisation (ICH) E9 (R1), published on November 20, 2019, provided 5 strategies to define the estimand: treatment policy, hypothetical, composite variable, while on treatment and principal stratum. In this article, we propose an estimand using a mix of strategies in handling ICEs. This estimand is an average of the "null" treatment difference for those with ICEs potentially related to safety and the treatment difference for the other patients if they would complete the assigned treatments. Two examples from clinical trials evaluating anti-diabetes treatments are provided to illustrate the estimation of this proposed estimand and to compare it with the estimates for estimands using hypothetical and treatment policy strategies in handling ICEs.

# 1. Introduction

The first randomized controlled trials (RCTs) in medicine were conducted as early as 1948.[1] RCTs have remained the gold standard for evaluation of the efficacy and safety of investigational interventions (drugs, biologics, devices, and other treatments). Ideally, all patients who participate in RCTs can adhere to the assigned treatment and complete the studies, allowing for the analysis of complete data. Often, however, post-baseline factors prevent patients from completing the designed treatment. Such factors include unsatisfactory efficacy, adverse events (AEs), or other reasons not related to the efficacy and safety of the treatment. These events, called *intercurrent events* (ICEs), either result in non-compliance with the study intervention (e.g., reduced dose level or discontinuation of the intervention), and/or use of additional medications for unsatisfactory efficacy (often called *rescue medication*). Therefore, ICEs, which affect the existence of observations or observed values that do not objectively reflect the originally intended treatment, create ambiguity in the estimation of the treatment effect.

The National Research Council (United States) Panel recommended that the estimand should be clearly defined before constructing the estimator.[2] The International Council for Harmonisation (ICH) E9 (R1)[3] further clarifies the process of defining an estimand by considering:

- Treatment(s) of interest – the treatment condition of interest and, as appropriate, the alternative treatment condition to which comparison will be made
- Handling of relevant ICEs with appropriate strategies. Five strategies are proposed: treatment policy, hypothetical, composite variable, while on treatment and principal stratum
- Population of interest
- Outcome variable (endpoint) at patient level
- Population-level summary of treatment effect

Although people generally define the estimand based on only one of these 5 strategies in handling ICEs, these strategies do not need to be mutually exclusive. Further discussions on choosing and defining the estimand and aligning the estimators with the estimands were recently published.[4,5,6] These publications further clarify (1) the need to evaluate what data are to be included, and (2) the separation of the definition of the estimand and handling of missing values.

An estimand for which ICEs are addressed using the treatment policy strategy and characterizes the treatment effect regardless of ICEs is generally preferred by many regulatory agencies. We call this estimand the *de facto estimand* in this article. This estimand is plausible if the treatment of interest is the study medication plus patients' choice of additional rescue medication and choice of taking or stopping the study medication. People who prefer this estimand generally argue it is the closest to the intent-to-treat (ITT) principle[7] and/or it mimics the treatment regimen in clinical practice and answers the question for treatment effectiveness.[8] ICH E9 (R1) has discussed the drawbacks of blindly following the ITT principle. In addition, the purpose of most randomized clinical trials is to assess the efficacy and safety of the desired treatment in a controlled setting that could be considerably different from real-world clinical practice (e.g., frequencies of visit schedules, allowed concomitant medications, etc.), so the assessment of the effectiveness in the real world may not be achievable regardless of how the estimand is defined.

Another widely used estimand uses a hypothetical strategy to handle ICEs. This estimand is the treatment difference as if all patients had completed the assigned treatment without ICEs. We call this hypothetical strategy the *typical* hypothetical strategy. However, it may not be convincing to consider whether a patient would have the benefit of the treatment when this patient could not tolerate the treatment. Lipkovich et al[9] propose another hypothetical strategy assuming patients with ICEs have the same response as if they had taken the control treatment.

The composite strategy is commonly used to address ICEs by incorporating patients with ICEs as non-responders in the definition of the endpoint (i.e., treatment failure). This addresses some of the drawbacks of the hypothetical strategy; however, handling ICEs only using the composite strategy does not consider the reasons for ICEs and may not be clinically meaningful for some cases. For example, while it is reasonable to consider those with treatment discontinuation due to AEs as non-responders, it may not be clinically meaningful to assume that those with treatment discontinuation due to reasons unrelated to efficacy or safety are non-responders.

The estimands for which ICEs are addressed using the principal stratum strategy are clinically important but generally difficult to estimate because the principal stratum cannot be observed in parallel-group studies.[10-15] A principal stratum is generally a hypothetical population of interest defined by the potential outcomes of a post-randomization variable. For example, one may be interested in the principal stratum that patients can adhere to both the experimental and control treatments, while only the adherence status for their assigned treatment can be observed.

In our experience, up to now, the primary estimand in a clinical trial has been almost exclusively defined using a single strategy in handling ICEs, regardless of the nature of ICEs. ICH E9 (R1) does not require all ICEs be handled the same way, as it may not be plausible to handle the treatment discontinuation due to adverse events in the same way as use of rescue medication. Akacha et al[16] propose tripartite estimands: the difference in the proportion of patients with treatment discontinuations due to AEs, the difference in the proportion of patients with treatment disconsolation due to lack of efficacy (LoE), and the treatment difference in the primary outcome for adherers. While the authors emphasize the importance to differentiate the reasons for treatment discontinuations and classify them into three categories: due to AE, due to LoE or due to administrative reasons, their interest in the treatment difference is for adherers (a principal stratum).

In this article, we introduce an *estimand* with a mix of strategies in handling ICEs according to the nature of individual ICEs. Data from 2 clinical trials evaluating anti-diabetes treatments (one superiority study and one non-inferiority study) are analyzed to illustrate the estimation for this new estimand.

## 2. Methods

The purpose of most RCTs is to assess the efficacy and safety of an experimental treatment under the ideal situation (i.e., without non-compliance, without dropout, and without rescue medication). However, it is often not possible to achieve the ideal situation due to ICEs. Therefore, we consider two questions of interest to determine the efficacy of an intervention for the population of all randomized patients, depending on people's viewpoints. First, what is the average efficacy in the ideal situation when all patients would take the study medication as randomized? This corresponds to the estimand using a hypothetical strategy to handle ICEs, i.e., using the potential outcome assuming patients would complete the assigned treatment regardless of ICEs. In the rest of this article, we call this estimand the *theoretic estimand*. Most past research on handling missing values actually intends to estimate this estimand. However, when an ICE occurs due to clinically insurmountable reasons, it may be argued that it is unreasonable to assume patients would have had good efficacy if they had completed the intervention. In other words, the hypothetical efficacy for those patients who are not able to complete the desired treatment can never be realized, despite the best effort of patients and investigators. Therefore, the second question is, what is the average efficacy for all randomized patients if those patients who are unable to complete the intervention due to insurmountable reasons (such as death or AEs) are considered to have no ("null") efficacy (i.e., the treatment effect under the null hypothesis)? We define a "hybrid" estimand , which uses a mix of hypothetical strategies according to the nature of ICEs (assuming the potential outcome is "null" [no effect] for those discontinued due to insurmountable reasons and as if patients complete the treatment for those with other intercurrent events). In the rest of this article, we call the estimand with this specific mix of strategies handling ICEs the *hybrid estimand*.

We classify all the ICEs into 3 categories:

- Category 1, including ICEs potentially related to safety
- Category 2, including ICEs potentially due to LoE
- Category 3, including ICEs due to administrative reasons

We consider the study medication has a relatively acute effect. For example, the response to the treatment should be apparent within a few weeks of taking the medication and should diminish after a few weeks of stopping the medication. Let $Z$ be the treatment indicator ($Z = 0$ for the control treatment, and $Z = 1$ for the experimental treatment), $Y(Z)$ denote the potential outcome (a continuous variable) for treatment $Z$, and $S(Z)$ denote the potential indicator whether a patient experiences an ICE of Category 1 when treated with treatment $Z$. Given an estimand, assume the null and alternative hypotheses are

$$H_0: \mu = \delta \quad \text{vs.} \quad H_a: \mu < \delta,$$

where $\mu$ is the treatment difference (estimand), and $\delta = 0$ for superiority studies and $\delta$ is the non-inferiority margin (NIM) for non-inferiority studies.

The *hybrid estimand* can be defined mathematically as

$$\begin{aligned}
\mu_h &= E\big\{\big[Y(1)\big(1 - S(1)\big) + (Y(0) + \delta)S(1)\big] - Y(0)\big\} \\
&= E\big\{[Y(1) - Y(0)]\big(1 - S(1)\big) + \delta \cdot S(1)\big\} \\
&= E[Y(1) - Y(0)|S(1) = 0] \cdot \Pr(S(1) = 0) + \delta \cdot \Pr(S(1) = 1) \quad (1)
\end{aligned}$$

Note $\mu_h \neq \{\Pr(S(1) = 0) \cdot E[Y(1)] + \Pr(S(1) = 1) \cdot (\delta + E[Y(0)])\} - E[Y(0)]$. An estimator for $\mu_h$ can be constructed as

$$\hat{\mu}_h = \frac{1}{n_1} \sum_{j=1}^{n_1} \left[ (1 - S_{1j})\hat{Y}_{1j}(1) + S_{1j}(\hat{Y}_{1j}(0) + \delta) \right] - \frac{1}{n_0} \sum_{j=1}^{n_0} \hat{Y}_{0j}(0), \qquad (2)$$

where $S_{1j}$ is the observed indicator for patient $j$ with an ICE of Category 1 for the experimental treatment group, $\hat{Y}_{zj}(z^*)$ is the estimated response for patient $j$ in treatment group $Z = z$ while treated with treatment $Z = z^*$. The estimation $\hat{Y}_{1j}(0)$ requires the framework of the principal stratification.

- For those patients with non-missing response in the absence of ICEs, the estimated value $\hat{Y}_{zj}(z)$ equals the observed value $Y_{zj}$.

- For patients who make a choice regarding treatment that represents an ICE of Categories 2 and 3 (i.e., due to LoE or due to administrative reasons), the responses after intercurrent events (if there are any) are treated as missing values and are imputed as if patients had completed the treatment. In this article, missing values as a result of Category 3 of ICEs were considered missing completely at random (MCAR) or missing at random (MAR); the missing data as a result of Category 2 of ICEs were considered MAR since the efficacy measurements were collected right before the ICEs. Alternative assumptions for missing values may also be considered, but we do not discuss them here.

- For patients with insurmountable ICEs (i.e., Category 1 of ICEs), the response is considered to be "null" (no effect). Specifically, responses (if there are any) after the ICEs are considered missing. For superiority studies, the response for patients with insurmountable intercurrent events in both treatment groups is considered to be like the control treatment. Therefore, the response for both treatments can be imputed using the observed data for patients in the control treatment group using a reference-based imputation.[8, 17-23] For non-inferiority studies, the NIM can be added (assuming smaller outcome means better efficacy) to the imputed response for patients in the experimental treatment group to make the treatment difference have a "null" effect.[24] Note this type of reference-based imputation assumes the data used for imputation can define the principal stratum $\{j: S_{zj}(1) = 1\}$. For example, if only the baseline variables in the experimental treatment arm are used for the imputations, we assume this stratum can be determined through a stochastic model only depending on baseline covariates but not on the post-baseline outcomes. This type of assumption is required for all reference-based imputation methods, but surprisingly, it is rarely explicitly pointed out.

## 3. Examples

We compare the estimators for three estimands listed in Table 1 for the change in Hemoglobin A1C (HbA1c) from baseline to the primary time point using data from two clinical studies (AWARD-1 [ClinicalTrials.gov Identifier: NCT01064687] and IMAGINE-3 [ClinicalTrials.gov Identifier: NCT01454284]):

- The estimator for the theoretic estimand is based on the mixed models with repeated measures (MMRM) analysis. The responses after intercurrent events were set to be missing before performing the analysis. The MMRM approach, which assumes MCAR or MAR, is widely used in analyzing data from clinical trials.

- The estimator for the de facto estimand included all observed data, used the jump-to-reference (J2R) approach[22] to impute missing values in the experimental treatment arm(s), and then applied MMRM to analyze the data. For non-inferiority studies, a NIM was added to the J2R imputed values.

- The estimator for the hybrid estimand treated the responses after ICEs as missing values, imputed missing values as a result of ICEs due to AEs using J2R, and then analyzed the data using MMRM. Similarly, for non-inferiority studies, a NIM was added to the J2R imputed values. In this analysis, the missing data due to other preventable ICEs were implicitly imputed with MMRM.

Note the above 3 estimators cannot be directly compared because they estimate different estimands. We perform the 3 analyses to illustrate the potential difference between estimands, realizing that the estimates are just one realization of the estimands. In both studies, the reasons for treatment discontinuation were collected in the clinical trial. However, some categories were vague, such as "investigator's decision", "subject's decision", etc. We reviewed the reasons for discontinuation, efficacy data (HbA1c and fasting serum glucose) and adverse events, and classified the ICEs into the 3 categories discussed in Section 2. More details can be found in Qu et al.[15, 16] When there was more than one potential reason for ICEs, we classified the ICEs based on the priority order of Categories 1, 2 and 3.

## 3.1. AWARD-1 Study

AWARD-1 is a randomized, parallel-arm study comparing dulaglutide 0.75 and 1.5 mg with exenatide twice daily and placebo.[25] There were 978 patients randomized to dulaglutide 0.75 mg, dulaglutide 1.5 mg, exenatide, or placebo with a 2:2:2:1 ratio. In this analysis, we compared dulaglutide 0.75 and 1.5 mg with placebo for the change in HbA1c from baseline to 26 weeks to show dulaglutide is superior to placebo.

Table 2 summarizes the proportion of patients with each category of ICEs. A considerably larger proportion of patients in the placebo arm had Category-2 ICEs compared to dulaglutide treatment arms.

The numbers of patients in each of the 3 arms and the estimates based on the 3 methods are shown in Table 3. The de facto estimand is consistent with the current FDA's request to ignore the use of rescue medication. The estimator for the de facto estimand provided the most conservative estimates for the treatment difference, while the estimators for the theoretic estimand and hybrid estimand provided similar estimates. The smaller estimated treatment effect for the the de facto estimand was primarily due to the higher reduction in the mean change in HbA1c for the placebo arm with the inclusion of the data collected after initiation of rescue medications. The estimated mean change in HbA1c from baseline to 26 weeks for the placebo arm was 0.64% when using the treatment policy strategy to handle ICEs and -0.41% when using the typical hypothetical strategy and the mix of hypothetical strategies.

## 3.2. IMAGINE-3 Study

IMAGINE-3 is a double blind, randomized study comparing 2 basal insulins.[26] There were 1,114 patients randomized to insulin peglispro or insulin glargine with a 3:2 randomization ratio. The primary objective was to show insulin peglispro was non-inferior to insulin glargine in the HbA1c level at 52 weeks using a non-inferiority margin of 0.4%. In this article, we considered a more commonly used endpoint: the change in HbA1c from baseline to 52 weeks. Note in linear models with adjustment for baseline HbA1c, the estimate for the treatment difference in the change in HbA1c from baseline to 52 weeks is the same as for HbA1c at 52 weeks. We analyzed the data as a non-inferiority study using a non-inferiority margin of 0.4% as specified in the original protocol.

Table 4 summarizes the proportion of patients with each category of ICEs. A larger proportion of patients with Category-1 ICEs in the insulin peglispro group compared to insulin glargine group. Table 5 shows the estimated mean change from baseline for each treatment arm and the treatment difference. The estimated mean treatment differences for the change in HbA1c from baseline to 52 weeks were -0.22%, -0.13%, and -0.17% for the theoretic, de facto and hybrid estimands. The estimates for the mean change from baseline to 52 weeks for insulin glargine were similar, and the major differences came from the estimates for insulin peglispro. Again, the treatment policy strategy provided the most conservative estimate. This is probably because in the estimation of the de facto estimand all missing values were imputed with a conservative approach under the "null" scenario and the response (if available) after stopping the study medication using "standard of care" basal insulin (mostly insulin glargine) was used, which in both cases might underestimate the benefit of the experimental treatment.

## 4. Summary and Discussions

Estimands and handling missing data are important topics in analyzing clinical trial data. ICH E9 (R1) provides guidance on the steps and principles to define the estimand. However, debate continues on the definition or selection of estimands, especially on which strategies to be used in handling ICEs. For example, different stakeholders (sponsors, regulatory agencies, payers and journal reviewers) may not agree on the strategy of handling ICEs. Each stakeholder may provide an example of the plausibility of their preferred strategy. For example, sponsors argue the hypothetical strategy should be used to handle the ICE of rescue medication use as the outcomes collected after the use of rescue medication may not reflect the treatment of interest. On the other hand, a regulatory agency may argue the hypothetical strategy is not plausible for handling ICE of AEs when assuming patients who cannot tolerate the treatment still have the full benefit of the treatment. To meet the requirement for different stakeholders, different estimands are often defined. For example, two estimands were used in PIONEER-2 Study: "the treatment policy" estimand (using the treatment policy strategy for all ICEs) and the "trial product" estimand (using the typical hypothetical strategy for all ICEs).[27]

In this article, we propose an estimand with a mix of strategies in handling ICEs, which defines the treatment difference as the average of the treatment difference for those who could adhere to the treatment and the "null" treatment difference (no effect) for those who could NOT adhere to the treatment. Those who could adhere to the treatment include patients who adhere to treatment or had ICEs potentially related to LoE or administrative reasons. Those who could NOT adhere to treatment include patients with ICEs potentially related to AEs. This estimand answers an important clinical question, "What is the average treatment effect when considering patients with ICEs due to AEs as receiving no benefit?" In the estimation of this new estimand, we recommend the data collected after ICEs should NOT be included in the analysis, and the responses for these patients should be treated as missing. In the examples, we imputed the missing values as a result of ICEs due to lack of efficacy or due to administrative reasons implicitly in a MMRM model under the assumption of MAR, and imputed missing values as a result of ICEs due to AEs using a reference-based imputation under the assumption of missing not at random (MNAR). Other assumptions of missingness may be used for the estimation for the estimand with a mix of strategies in handling ICEs, but we advocate the underlying reasons for the missing values should be taken into consideration in determining the mechanisms of missingness, as recommended in literature. [28, 29] Therefore, considering the nature of ICEs is not only important for defining the estimand, but also for making the appropriate assumption in imputing the resulting missing values.

Jump-to-reference imputation imputes the potential outcome for patients with ICEs of AE assuming "similar" patients in the control arm can be identified for imputation (i.e., the identification of the principal stratum of patients with ICEs due to AE or consistently estimating the probability of belonging to the principal stratum in the control arm). The probability for a patient belonging to the principal stratum can be estimated if the potential ICE under one treatment is independent of the potential outcome under the other treatment (*principal ignorability* assumption) [30], or under weaker but more complex assumptions[14]. While estimating the potential outcome for a principal stratum may be generally challenging, there are two reasons that it is less of a concern in the two examples to which we applied it. First, the principal stratum variable (ICE due to AE) is generally considered to be independent of the primary outcome of HbA1c for the diabetes treatment. Second, the principal stratum size is relatively small (approximately 10-20% of the overall population). Note even for the estimand using the treatment policy strategy to handle ICEs, a jump-to-reference imputation for missing values also requires the same "principal stratum" assumption, although most of time the assumption is not explicitly stated.

We summarized 3 estimands in two clinical trials for evaluating anti-diabetes treatment: the theoretic estimand using the typical hypothetical strategy to handle all ICEs, the de facto estimand using the treatment policy strategy to handle all ICEs, and the hybrid estimand using a mix of hypothetical strategies in handling ICEs according to the nature of ICEs. In both examples, the estimates for the de facto estimand seemed to be the most conservative, and the estimates for the theoretic estimand based on the assumption of MAR provided the most liberal estimate for the treatment difference. The hybrid estimand, which has a clear and reasonable clinical meaning, addresses some criticisms for both theoretic and de facto estimands and can be a reasonable option for the primary estimand to evaluate treatment efficacy in clinical trials.

We applied this framework for the primary objective, but it can certainly be applied to supplemental estimands and/or the sensitivity analysis for the primary/secondary objectives (with different assumptions for imputation). For example, with the same estimand, one can introduce a sensitivity parameter to impute the missing values due to ICEs related to AE.[31,32] Note it is possible that one analysis could serve as the primary analysis for one estimand and a sensitivity analysis for another estimand.

In the two examples, we used *one* type of hybrid strategies in handling ICEs. There are other possible mixes of strategies in defining estimands. For example, when the outcome is a binary variable (e.g., whether patients respond to the treatments), a composite strategy can be used to handle ICEs due to LoE and treat these patients as non-responders if the clinical interest is that patients who discontinued treatment due to LoE or use of rescue medication are considered not meeting the treatment objective. Note the hypothetical outcome for patients with ICEs related to AE may not have to be "null" effect. The "no benefit" potential outcome is no improvement in the outcome from baseline. If using this hypothetical strategy to handle ICEs, the potential outcome can be imputed using a return-to-baseline imputation.[33] The "no benefit" potential outcome can also be considered to be the outcome assuming patients complete the study but there is no treatment from the occurrence of ICEs to the study end. If using this hypothetical strategy in handling ICEs, the outcomes for patients who have ICEs but complete the study may be used in the analysis, and for patients who have ICEs and discontinue the study, the potential outcome can be imputed using patients with similar ICEs but with observed outcomes (retrieved dropout imputation).[34] Each of the above hypothetical strategies requires certain assumptions. Some of above hypothetical strategies may be especially useful for non-inferiority studies in which different hypothetical strategies may be used to define "no benefit". In this article, we use the "null hypothesis" hypothetical strategy to handle the ICEs related to AE, for which the estimand is equal to the treatment effect under the null hypothesis.

To estimate the hybrid estimand, the key is accurate collection of the reasons for treatment discontinuation. This should not be an extra burden for data collection, as the exact reasons for treatment discontinuation should be an expectation in clinical trials even if we do not use the mix of strategies in handling ICEs. This may require more detailed information for the reasons for treatment discontinuation, and may allow for multiple categories of reasons (e.g., AE and LoE). In addition, we should classify the ICEs proactively before unblinding the treatment code to prevent potential bias. Finally, the proportion of patients with each category of ICEs may be compared between treatments to assess the objectivity of classification. For example, if Category-3 ICEs are not similar between treatments for double-blind studies, the classification may need further investigation.

**Acknowledgement**

We thank Drs. Ilya Lipkovich, Dawn Brooks and Angelyn Bethel for their scientific review of this manuscript and valuable comments, and Dr. Brad Woodward for useful suggestions. We would also like to thank the two anonymous referees for their valuable comments that helped significantly improve the presentation of this article.

Available at https://www.ema.europa.eu/en/documents/scientific-guideline/guideline-missing-data-confirmatory-clinical-trials_en.pdf.

Table 1. Three estimators for certain combinations of estimands, analysis datasets, and methods of handling missing values.

| Estimand | Analysis Dataset | Estimator and Handling Missing Values |
|---|---|---|
| The theoretic estimand | All data on treatment before ICEs | MMRM based on MAR assumption |
| The de facto estimand | All available data, regardless of ICEs | J2R MI for all missing values (MNAR); for non-inferiority studies, the NIM is added to the imputed values<br>MMRM on complete data |
| The hybrid estimand | All data on treatment before ICEs | J2R MI for missing values for patients with Category 1 ICEs (MNAR); for non-inferiority studies, the NIM is added to the imputed values<br>MMRM including imputed data and assuming MAR for other missing data |

AE, adverse event; J2R, jump-to-reference; MAR, missing at random; MI, mixed imputation; MMRM, mixed models with repeated measures; NIM, non-inferiority margin; ICE, intercurrent events.

Table 2. Summary of ICEs in the AWARD-1 study (first 26 weeks)

| ICE Category* | Placebo (N=141) | Dulaglutide 0.75 mg (N=280) | Dulaglutide 1.5 mg (N=279) |
|---|---|---|---|
| Patients with ICEs | 36 (25.5%) | 33 (11.8%) | 27 (9.7%) |
| Category 1 | 9 (6.4%) | 18 (6.4%) | 15 (5.4%) |
| Category 2 | 23 (16.3%) | 10 (3.6%) | 4 (1.4%) |
| Category 3 | 4 (2.8%) | 5 (1.8%) | 8 (2.9%) |

*Category 1 includes TDC due to AE, and TDC due to other reasons but with at least one persistent AE reported before DC; Category 2 includes TDC due to LoE, use of rescue med, and TDC due to other reasons with no persistent AEs reported before TDC and with deteriorated efficacy prior to TDC; Category 3 includes all other ICEs.

Abbreviations: AE, adverse events; ICE, intercurrent events; LoE, lack of efficacy; TDC, treatment discontinuation.

Table 3. Estimated mean change in HbA1c from baseline to 26 weeks and treatment differences compared to placebo for the AWARD-1 study

| Method | Placebo (N=141) Mean (SE) | Dulaglutide 0.75 mg (N=280) Mean (SE) | Dulaglutide 1.5 mg (N=279) Mean (SE) | Dulaglutide 0.75 mg vs. Placebo Mean (95% CI) | Dulaglutide 1.5 mg vs. Placebo Mean (95% CI) |
|---|---|---|---|---|---|
| (A) MMRM for the theoretic estimand | -0.41 (0.07) | -1.26 (0.05) | -1.51 (0.05) | -0.85 (-1.02, -0.68) | -1.10 (-1.27, 0.93) |
| (B) Placebo-based imputation for the de facto estimand | -0.64 (0.07) | -1.29 (0.05) | -1.52 (0.05) | -0.65 (-0.82, -0.48) | -0.88 (-1.05, -0.71) |
| (C) Mixed imputation for the hybrid estimand | -0.41 (0.07) | -1.25 (0.05) | -1.51 (0.05) | -0.84 (-1.01, -0.67) | -1.10 (-1.27, -0.93) |

Abbreviations: CI, confidence interval; HbA1c, Hemoglobin A1c; MMRM, mixed models with repeated measures; SE, standard error.

Table 4. Summary of ICEs in the IMAGINE-3 study

| ICE Category* | Insulin Glargine (N=449) | Insulin Peglispro (N=663) |
|---|---|---|
| Patients with ICEs | 81 (18.0%) | 154 (23.2%) |
| Category 1 | 24 (5.3%) | 70 (10.6%) |
| Category 2 | 8 (1.8%) | 15 (2.3%) |
| Category 3 | 49 (10.9%) | 69 (10.4%) |

*Category 1 includes TDC due to AE, and TDC due to other reasons but with at least one persistent AE reported before DC; Category 2 includes TDC due to LoE, use of rescue med, and TDC due to other reasons with no persistent AEs reported before TDC and with deteriorated efficacy prior to TDC; Category 3 includes all other ICEs.

Abbreviations: AE, adverse events; ICE, intercurrent events; LoE, lack of efficacy; TDC, treatment discontinuation.

Table 5. Estimated mean change in HbA1c from baseline to 52 weeks and treatment differences for the IMAGINE-3 study

| Method | Insulin Glargine (N=444) Mean (SE) | Insulin Peglispro (N=648) Mean (SE) | Insulin Peglispro vs. Insulin Glargine Mean (95% CI) |
|---|---|---|---|
| (A) MMRM for the theoretic estimand | -0.24 (0.04) | -0.46 (0.03) | -0.22 (-0.32, -0.12) |
| (B) Placebo-based imputation for the de facto estimand | -0.26 (0.04) | -0.38 (0.03) | -0.13 (-0.22, -0.03) |
| (C) Mixed imputation for the hybrid estimand | -0.24 (0.04) | -0.41 (0.03) | -0.17 (-0.27, -0.07) |

Abbreviations: CI, confidence interval; HbA1c, Hemoglobin A1C; MMRM, mixed models with repeated measures; SE, standard error.